\documentclass[fleqn,twoside]{article}
\usepackage{espcrc2}
\usepackage{graphicx}

\newcommand{\beeq}{\begin{equation}}
\newcommand{\eneq}{\end{equation}}

\newcommand{\beeqn}{\[}
\newcommand{\eneqn}{\]}

\newcommand{\beeqa}{\begin{eqnarray*}}
\newcommand{\eneqa}{\end{eqnarray*}}

\title{
{\vspace{-1.2em} \parbox{\hsize}{\hbox to \hsize
{\hss  \normalsize TRINLAT-03/04 , UPRF-2003-19}}} \\
The $2+2$ anisotropic Wilson gluon action with applications}
\author{Giuseppe~Burgio\address[TCD]{The TrinLat Collaboration \\ 
        School of Mathematics, Trinity College, Dublin 2, Ireland},
        Alessandra~Feo\addressmark[TCD]\address[parma]{Dipartimento di Fisica, Universit\`a di Parma and 
INFN Gruppo Collegato di Parma, Parco Area delle Scienze, 7/A, 43100 Parma, Italy}
       \thanks{Talk given by Alessandra~Feo},
        Mike~Peardon\addressmark[TCD]
        and
        Sin\'ead~M.~Ryan\addressmark[TCD]}
       
\begin{document}
\begin{abstract}
A generalization of anisotropic lattices to include a 2+2 discretization is discussed.
As a part of this program, we determine the one-loop correction to the gluon self-energy and analyse 
the restoration of Lorentz invariance in on-shell states for both the 2+2 and 3+1 cases. 
A 2+2 Wilson-like fermion action is also considered.
\end{abstract}
\vspace{1pc}
\maketitle

\section{MOTIVATION}

The usual approach for anisotropic lattices is 3+1 which makes the temporal direction fine while 
keeping the spatial directions relatively coarse. 3+1 has been successful in the study of 
glueball states~\cite{peardon} and heavy hadrons (see references in~\cite{ryan}).
In a 2+2 discretization a temporal and a spatial direction is made fine while the other two 
spatial directions remain coarse. The motivation for this choice is to calculate decays which produce 
a high-momentum daughter particle, where the momenta grows quickly 
in lattice calculations. Decay examples are $B \to \pi\ell\nu $ or $B \to K^* \gamma$. 
We hope that making one spatial direction fine and injecting all momentum along that direction will 
keep the discretization errors of order ${\cal O}(ap)$ 
small for high momenta thereby extending the range of momentum accesible to lattice calculations.

\section{DIFFERENCES BETWEEN 3+1 AND 2+2}

For the 3+1 case the physics of the Euclidean invariant Yang-Mills theory is reproduced, provided
that $\xi= a_c/a_f$ takes the desired value, while for the 2+2 case, the recovery of a Euclidean invariant 
continuum limit theory is not guaranteed since there are two free parameters and a single ratio
of scales. In order to ensure the recovery of the Euclidean invariance in the continuum limit 
the relative weights between the three operators must be determined to ensure Lorentz
invariance in on-shell Green functions. This fine tuning must be fixed, either non-perturbatively
\cite{burgio} or perturbatively.

We focus here on the perturbative approach. 
We follow the notation of Ref.~\cite{trinlat} and first concentrate in the pure gluon sector of 
the $SU(N_c)$ Yang-Mills theory. The action reads,
\beeq
S_{\rm g} = \beta \hspace{-0.1 cm} \sum_{x,\mu,\nu} c_{\mu \nu} [ 1 - \frac{1}{2 N_c} {\rm Tr} \, 
(P_{\mu \nu}(x) + P^\dagger_{\mu \nu}(x) ) ]
\eneq
where the one-loop coefficients, $c_{\mu \nu}$ can be parametrized as 
\beeq
              c^{(2+2)}_{\mu\nu}= \left\{ \begin{array}{lll}
              \xi^2(1+ \eta_{ff}^{(1)} g^2 +O(g^4)) & \, \mbox{f-f} \\
              1+\eta_{cf}^{(1)} g^2 +O(g^4)         & \, \mbox{c-f} \\
              \frac{1}{\xi^2}(1+\eta_{cc}^{(1)} g^2 +O(g^4))& \, \mbox{c-c}\end{array} \right. 
\eneq
for 2+2 and 
\beeq
              c^{(3+1)}_{\mu\nu}= \left\{ \begin{array}{lll}
              \xi (1+\eta_{cf}^{(1)} g^2 +O(g^4)) & \, \mbox{c-f} \\
              \frac{1}{\xi}(1+\eta_{cc}^{(1)} g^2 +O(g^4))& \, \mbox{c-c}\end{array} \right. 
\eneq
for 3+1. These three parameters must be tuned in order to restore the on-shell Lorentz invariance.
At tree-level it is easy to demonstrate that 
$c_{cc} = \frac{1}{\xi^2}, c_{cf} = 1, c_{ff} = \xi^2$ for 2+2 and
$c_{cc} = \frac{1}{\xi}, c_{cf} = \xi$ for 3+1. The determination of the one-loop
coefficients $\eta_{ff}^{(1)}, \eta_{cf}^{(1)}, \eta_{cc}^{(1)}$ requires more work \cite{trinlat}
and an algebraic manipulation in order to deal with the plethora of terms appearing during the calculation 
is required (the core of the code is similar to that used in Ref. \cite{lpt}).

To get the Feynman rules for a general anisotropic Wilson action, the matching of the gluon action
with its continuum counterpart is made in terms of dimensionless fields, $U_\mu(n) = e^{i \phi^b_{\mu}(n) T^b}$. 
The dimensionful field $A^b_{\mu}(x)$ is reintroduced by $\phi_{\mu}^b(n) = g a \xi_{\mu} A^b_{\mu}(x)$
where $\xi_{\mu} = 1 $ for $\mu$ fine and $\xi_{\mu} = \xi $ for $\mu$ coarse.
We find a general rule: for any vertex which has a continuum analogue the form, in terms of the anisotropic 
dimensionful momenta, resembles the isotropic case since the anisotropy lies only in the different momenta cut-off
\cite{trinlat}.

The one-loop correction to the action coming from $c_{\mu \nu}$ leads to an extra vertex which reads 
\beeqn
-(2 \pi)^4 \delta^4 (k+k') g^2  \delta^{ab}  (\delta_{\mu\nu} \sum_{\rho} \eta^{(1)}_{\mu\rho}
	\hat{k}_{\rho}^2-\eta^{(1)}_{\mu\nu}\hat{k}_{\mu}\hat{k}_{\nu})
\eneqn
where $ \eta^{(1)}_{\mu\nu} = \eta_{f\!f}^{(1)}$ for $\mu,\nu = $ fine, 
$ \eta^{(1)}_{\mu\nu} = \eta_{c f}^{(1)}$ for $\mu= $ fine(coarse) and $\nu= $ coarse(fine) and
$ \eta^{(1)}_{\mu\nu} = \eta_{c\!c}^{(1)}$ for $\mu,\nu = $ coarse;
while for the 3+1 case one can set $\eta_{f\!f}^{(1)}=0$.

\section{RESULTS}

The calculation of the one-loop correction to the gluon self-energy involves five Feynman diagrams.
Each diagram is a function of the external momenta $p$
\beeq
G(p) = \int {d^4 k\over (2 \pi)^4} F(k,p) \, ,
\label{int}
\eneq
where $k$ is the integration momenta. Since we are interested in the study of the continuum limit 
of Eq.~(\ref{int}) and the propagators are massless, our procedure of anisotropic renormalization 
which requires a Taylor expansion around $p=0$ gives rise to infrared divergences. One way to deal with the 
problem is to introduce a mass term, $m$. The integral can then be separated in two parts which are both
separately divergent for $m \to 0$ but the divergences cancel when the two contributions are summed.

The one-loop corrections to the gluon self-energy using the Feynman rules for a general anisotropy 
do not satisfy Lorentz invariance. As previously mentioned, the three parameters in the action must 
be tuned in order to restore the symmetry.
Other Lorentz-breaking terms such as $\delta_{\mu \nu}/a^2 $ and $\delta_{\mu \nu} p_\mu^2 $ are independent
of $\xi$ and cancel when all the contributions are summed (as in the isotropic case). 
This is a first non-trivial check of our methodology. Another check is the possibility to easily change 
from the 2+2 case to the 3+1 case (the only difference are the numerical integrals used, once the external 
momenta in Eq.~(\ref{int}) have been extracted). This makes our procedure very powerful and 
also allows us to check our general anisotropic results with 3+1 results already in the literature.
The remaining Lorentz-breaking artifacts are calculated by imposing that the physical eigenvalues of the
gluon propagator vanish at $E^2 = p^2$.

The one-loop coefficients for the Lorentz restoration are
\beeqa
\eta_{cc}^{(1)} - \eta_{cf}^{(1)} &=&
		-\frac{1}{2 N_c}\Bigg[ {\cal B}_\xi^c(1,1)-\frac{1}{4}\Bigg]+  \nonumber \\ 
&& \hspace{-2 cm} N_c\Bigg[ -\frac{1}{16} + \frac{{\cal B}_\xi(1)}{6}\bigg( \frac{7}{2} + \frac{1}{\xi^2} \bigg) +
			\frac{{\cal B}_\xi^c(1,1)}{4} - \nonumber \\ 
&& \hspace{-2.5 cm} \frac{{\cal B}_\xi^c(2,1)}{3} \bigg(2 \!+\! \frac{5}{2\xi^4} 
                    \!+\!  \frac{11}{2 \xi^2} \bigg) 
      + \frac{{\cal B}_\xi^f(2,1,1)}{6} \bigg(\frac{1}{2} \!+\! \xi^2 \bigg) \Bigg] 
\eneqa
and 
\beeqa
\eta_{f\!f}^{(1)} - \eta_{cf}^{(1)} &=&
		\frac{1}{2 N_c} \bigg[ \frac{1}{4}-\frac{1}{2 \xi^2} +\frac{{\cal B}_\xi^c(1,1)}{\xi^4}\bigg] 
   - \nonumber \\
&& \hspace{-2.5 cm} \frac{N_c}{2} \bigg[ \frac{1}{4} \bigg( \frac{1}{2} \!-\! \frac{1}{\xi^2}\bigg) + 
			{\cal B}_\xi(1) \bigg(\frac{1}{2} \!+\! \frac{1}{3 \xi^2} \bigg) 
   \!+\!  \frac{{\cal B}_\xi^c(1,1)}{2 \xi^4} - \nonumber \\    
   &&  \hspace{-2.5 cm} \frac{{\cal B}_\xi^c(1,1)}{2 \xi^4} - 
			{\cal B}_\xi^c(2,1) \bigg( \frac{5}{3 \xi^4} \!+\! \frac{1}{\xi^2} \!+\! \frac{1}{6} \bigg)
   \!-\! \frac{{\cal B}_\xi^f(2,1,1)}{3 \xi^2} \bigg] 
\eneqa
for 2+2 while 
\beeqa
\eta_{cc}^{(1)} - \eta_{cf}^{(1)} &=&
		\frac{N_c}{\xi} \Bigg[ \frac{{\cal B}_\xi(1)}{6} \bigg(\frac{\xi^2}{3} + \frac{19}{6} + 
		\frac{7}{2 \xi^2}\bigg) + \nonumber \\ 
&& \hspace{-2.5 cm} \frac{{\cal B}_\xi^c(1,1)}{4} \bigg(1 + \frac{1}{\xi^2} \bigg)  
   -{\cal B}_\xi^c(2,1) \bigg(\frac{1}{3}+\frac{11}{6 \xi^2} +\frac{5}{2 \xi^4} \bigg) 
		    \nonumber \\ 
&& \hspace{-2.5 cm} -\frac{1}{8}\Bigg]  - \frac{1}{2 \xi N_c} \Bigg[{\cal B}_\xi^c(1,1)\bigg(1+\frac{1}{\xi^2} 
		    \bigg) - \frac{1}{2} \Bigg]
\eneqa
for 3+1. 

\section{FULL THEORY}

An improved fermion action for 2+2 which contains an $O(a_c^3, a_f)$ discretization error is also presented.
For the fine components: $Z$ and $T$, we use the standard Wilson discretization with a 
nearest-neighbour representation of $\sum_{f=Z,T} \nabla_{f}^2$ to remove the doublers in these two directions,
\beeqa
\Delta^{(1)} \phi(x)&=& \frac{1}{2a} (\phi(x + a) - \phi(x - a)) \nonumber \\
\Delta^{(2)} \phi(x)&=& \frac{1}{a^2} (\phi(x + a) + \phi(x - a) 2 \phi(x)) 
\eneqa
while for the coarse components: $X$ and $Y$, we use a Hamber-Wu discretization with a 
next-to-nearest-neighbour representation of $\sum_{c=X,Y} \nabla_{c}^4$ to remove the doublers in the two
coarse directions. An improved discretization of the coarse axes terms in the Dirac operator,
$\sum_{c=X,Y} \gamma_c \partial_c $ is used, since the simplest one produces a discretization error of
${\cal O}(a_c^2)$,
\beeqa
\Delta^{(1,imp)} \phi(x)&=& \frac{1}{a} (\frac{2}{3} [\phi(x + a) - \phi(x - a)] - \nonumber \\ 
&& \frac{1}{12} [\phi(x + 2a) - \phi(x - 2a)]) \nonumber \\
\Delta^{(4)} \phi(x) &=& \frac{1}{a^4} ([\phi(x + 2a) - \phi(x - 2a)] - \nonumber \\ 
&& \hspace{-1 cm} 4 [\phi(x + a) - \phi(x - a)] + 6 \phi(x))  \, .
\eneqa
Making the anisotropic lattice Dirac operator gauge-covariant we have
\beeqa
S_q &=& \bar \psi \Bigg( m + \sum_{c=X,Y} (\gamma_c \Delta_c^{(1,imp)} + s a_c^3 \Delta_c^{(4)}) \\
&& + \sum_{f=Z,T} (\gamma_f \Delta_f^{(1)} + \frac{r a_f}{2} \Delta_f^{(2)} ) \Bigg) \psi
\eneqa
where $m$ is the bare mass of the fermion.

The more general anisotropic fermion action is
\beeq
S_q = \sum_{x,y} \psi(x) M_{x,y} \psi(y)
\eneq
 with 
\beeqa
&& \hspace{-0.6 cm} (M \psi)(x)= c_0 \delta_{xy} +
\nonumber \\
&& \sum_{f} \bigg[ ( c_{1f} + c_{2f} \gamma_f) U_f(x) \psi(x + \hat{f}) +
\nonumber \\
&& ~~~~~~~  ( c_{1f} - c_{2f} \gamma_f) U^{\dagger}_f(x - f) \psi(x - \hat{f})
   \bigg] +
\nonumber \\
&&  \sum_{c} \bigg[ ( c_{1c} + c_{2c} \gamma_c) U_c(x) \psi(x + \hat{c}) +
\nonumber \\
&&  ~~~~~~~ ( c_{1c} - c_{2c} \gamma_c) U^{\dagger}_c(x - \hat{c}) \psi(x - \hat{c}))  
   \bigg] +
\nonumber \\
&&  \sum_{c} \bigg[ ( c_{3c} + c_{4c} \gamma_c)  U_c(x) U_c(x + \hat{c}) \psi(x + 2 \hat{c}) +
\nonumber \\
&&  ( c_{3c} - c_{4c} \gamma_c)  U^{\dagger}_c(x - \hat{c}) U^{\dagger}_c(x - 2 \hat{c}) \psi(x - 2 \hat{c}) \bigg] \, .
\eneqa

The study of the fine tuning of this action is work in progress. First results of this action
applied to the 3+1 case are in Ref.~\cite{ryan2}.

\section{OUTLOOK}

A generalization of the Wilson discretization to an anisotropic lattice with two fine and two
coarse directions has been shown, in particular we have emphasized the feasibility of doing lattice 
perturbation theory on a general anisotropic lattice which shows the power of the method.

\section*{ACKNOWLEDGMENT}

This work was partially funded by the Enterprise-Ireland grants SC/2001/306 and 307.

\end{document}